 \documentclass[onecolumn,12pt,draft]{IEEEtran} 

\usepackage{epsf}

\begin{document}

\title{
On the ``SIR''s (``Signal''-to-``Interference''-Ratio) in Discrete-Time 
Autonomous Linear Networks 
} 

\author{Zekeriya Uykan \thanks{Z. Uykan is with Helsinki University of Technology, Control
Engineering Laboratory, FI-02015 HUT, Finland, and Nokia Siemens Networks, Espoo, Finland. 
E-mail: zekeriya.uykan@hut.fi, zuykan@seas.harvard.edu.
The author has been a visiting scientist at Harvard University Broadband Comm Lab., Cambridge, 02138, MA,
since September 2008
and this work has been performed during his stay at Harvard University.
} 
}


\maketitle
\begin{abstract}

In this letter, we improve the results in \cite{Uykan09b} 
by relaxing the symmetry assumption and also taking the noise term into account. 
The author examines two discrete-time autonomous linear systems 
whose motivation comes from a 
neural network point of view in \cite{Uykan09b}. Here, we examine the following discrete-time autonomous 
linear system: ${\mathbf x}(k+1) = {\mathbf A} {\mathbf x}(k) + {\mathbf b}$ where 
${\mathbf A}$ is any real square matrix with linearly independent eigenvectors 
whose largest eigenvalue is real and its norm is larger than 1, and 
vector ${\mathbf b}$ is constant. 
Using the same ``SIR'' (``Signal''-to-``Interference''-Ratio)  concept as in \cite{Uykan09a} and \cite{Uykan09b}, 
we show that the ultimate ``SIR'' is equal to $\frac{a_{ii}}{\lambda_{max} - a_{ii}}$, $i=1, 2, \dots, N$, where 
$N$ is the number of states, $a_{ii}$ is the diagonal elements of matrix ${\bf A}$, 
 and $\lambda_{max}$ is the (single or multiple) eigenvalue with maximum norm. 

\end{abstract}


\section{Introduction \label{Section:INTRO}}

\PARstart{T}{his} letter improves so-called "SIR" results in \cite{Uykan09b} where 
the author analyzes two discrete-time autonomous linear systems 
whose motivation comes from a neural network point of view.  
In \cite{Uykan08b} a dynamic-system-version of SIR ("Signal"-to-"Interference" Ratio) concept is 
introduced. Let's first present the definition of the SIR in \cite{Uykan09a}.

The SIR is an important entity in commucations engineering 
which indicates the quality of a link between a transmitter and a receiver in a multi transmitter-receiver 
environment (see e.g. \cite{Rappaport96}, among many others).  
For example, let $N$ represent the number of transmitters and receivers using the same channel. Then the received 
SIR at receiver $i$ is given by (see e.g. \cite{Rappaport96}) 

\begin{equation} \label{eq:cir}
SIR_i(k) = \gamma_i(k) = \frac{ g_{ii} p_i(k)}{ \nu_i + \sum_{j = 1, j \neq i}^{N} g_{ij} p_j(k) },  \quad i=1, \dots, N
\end{equation}

where  $p_i(k)$ is the transmission power of transmitter $i$ at time step $k$, $g_{ij}$ is
the link gain from transmitter $j$ to receiver $i$ (e.g. in case of cellular radio systems, 
$g_{ij}$ involves path loss, shadowing, etc) and $\nu_i$ represents the
receiver noise at receiver $i$. 
Typically, in cellular radio systems, every transmitter tries to optimize its 
power $p_i(k)$ such that the received SIR(k) (i.e., $\gamma_i(k)$) in eq.(\ref{eq:cir}) is kept at a 
target SIR value, $\gamma_i^{tgt}$. 

The author defines the following dynamic-system-version of the ``Signal-to-Interference-Ratio (SIR)'', 
denoted by $\theta_i(k)$, by rewriting the eq.(\ref{eq:cir}) with neural networks terminology 
in \cite{Uykan09a} and \cite{Uykan09b}: 

\begin{equation} \label{eq:cirA}
{\theta_i}(k) = \frac{ a_{ii} x_i(k)}{ b_i + \sum_{j = 1, j \neq i}^{N} a_{ij} x_j(k) },  
	\quad i=1, \dots, N
\end{equation}

where 
$\theta_i(k)$ is the defined fictitious ``SIR'' at time step $k$, 
$x_i(k)$ is the state of the $i$'th neuron, 
$a_{ii}$ is the feedback coefficient from its state to its input layer,  
$a_{ij}$ is the weight from the state of the $j$'th neuron to the input of the 
$j$'th neuron, and $b_i$ is constant.

In this paper, we improve the results in \cite{Uykan09b} where 
the author examines the ultimate SIR in the following two discrete-time autonomous linear systems 
whose motivation comes from a neural network point of view.  

\begin{enumerate}

\item  The discretized autonomous linear system:

\begin{equation} \label{eq:Diff_Linear}
{\mathbf x}(k+1) = \big( {\bf I} + \alpha ( -r {\bf I} + {\bf W} ) \big) {\mathbf x}(k)
\end{equation} 

which is obtained by discretizing the autonomous continuous-time
linear system in \cite{Uykan09a} using Euler method; where
${\bf I}$ is the identity matrix, $r$ is a positive real number, and
$\alpha >0$ is the step size. 

\item A more general linear system with symmetric weight matrix 

\begin{equation} \label{eq:generalSymmetric} 
{\mathbf x}(k+1) = ( -\rho {\bf I} + {\bf W}  ) {\mathbf x}(k) 
\end{equation} 

where ${\bf I}$ is the identity matrix, $\rho$ is a positive real number, and 
$(-\rho {\bf I} + {\bf W})$ is the system matrix.  

\end{enumerate}

In \cite{Uykan09b}, the above linear systems with symmetric matrices and with zero 
noise term are examined.  In this paper, we relax these two conditions, and 
examine the SIR in the general discrete-time autonomous linear system in eq.(\ref{eq:generalDLS}) 
where the weight matrix ${\mathbf A} \in R^{N \times N}$ is any real matrix whose largest 
(single or multiple) eigenvalue is real and its norm 
is larger than 1.

It's well-known that in the system of eq.(\ref{eq:generalDLS}), 
the eigenvalues of the weight (system) matrix solely determine the stability of the system.  
If the spectral radius of the matrix is larger than 1, then the system is unstable. 
The spectral radius of an $N \times N$ matrix is equal to $\max \{  |\lambda_i| \}_{i=1}^{N}$ 
where $\lambda_i$ is the eigenvalues of the matrix. In this 
paper, we examine the linear systems 
with system matrices whose spectral radius is larger than 1. 

The paper is organized as follows:  The results for system eq.(\ref{eq:generalDLS}) are 
presented in section \ref{Section:USIR}. Section \ref{Section:SimuResults} presents some 
simple simulation results, followed by the conclusions in 
Section \ref{Section:CONCLUSIONS}.

\vspace{0.2cm}

\section{``SIR'' in Discrete-Time Autonomous Linear Networks}
\label{Section:USIR}

In this paper, we examine the SIR in the following autonomous linear system 

\begin{equation} \label{eq:generalDLS}
{\mathbf x}(k+1) = {\mathbf A} {\mathbf x}(k) + {\mathbf b}
\end{equation}

where ${\mathbf x}(k) \in R^{N \times 1}$ is state vector at step $k$, 
${\mathbf A} \in R^{N \times N}$ is 
any real square matrix with linearly independent eigenvectors 
whose largest eigenvalue is real and its norm is larger than 1, 
and vector ${\mathbf b}$ is constant. In what follows, we present the main result of this paper: 

\vspace{0.2cm}
\emph{Proposition:}
\vspace{0.2cm}

In the discrete-time linear system of eq.(\ref{eq:generalDLS}), 
if the matrix ${\mathbf A}$ has linearly independent eigenvectors, and if the  
(single or multiple) eigenvalue with the 
greatest norm is real and its norm is greater than 1, then 
the defined "SIR" (${\theta_i}(k)$) in eq.(\ref{eq:cirA}) for any state $i$  
converges to the following constant within a finite step number denoted as 
$k_T$ for almost any initial vector ${\bf x}(0)$. 

\begin{equation} \label{eq:theta_const_general} 
\theta_i(k \geq k_T) =  \frac{ a_{ii} }{ \lambda_{max} - a_{ii} }, \quad \quad i=1,2, \dots, N 
\end{equation} 

where $\lambda_{max}$ is the (single or multiple) eigenvalue with maximum norm and 
$a_{ii}$ is the diagonal elements of ${\bf A}$.

\begin{proof} 

From eq.(\ref{eq:generalDLS}), 
\begin{equation} \label{eq:genDLsolution} 
{\mathbf x}(k) = {\bf A}^{k} {\mathbf x}(0) + \sum_{j=0}^{k-1} {\bf A}^{j} {\bf b} 
\end{equation} 

where ${\bf x}(0)$ shows the initial state vector at step zero. It's well known that any real 
square matrix ${\bf A}$ whose eigenvectors are linear independent can be decomposed into 
(see e.g. \cite{Bretscher05}) 

\begin{eqnarray} 
{\bf A} & = & {\bf V} {\bf D}_{\bf \lambda} {\bf V}^{-1}  \label{eq:symW2} \\ 
{\bf A} & = & {\bf V} {\bf D}_{\bf \lambda} {\bf U} \label{eq:symW3} 
\end{eqnarray} 

where the matrix ${\bf V}$ has the eigenvectors in its columns, and 
matrix ${\bf D}_{\bf \lambda}$ has the eigenvalues in its diagonal, and where 

\begin{equation} \label{eq:VU-1} 
{\bf U} = {\bf V}^{-1} 
\end{equation} 

The matrices in eq.(\ref{eq:symW3}) are given as   

\begin{equation} \label{eq:matMVDU}
{\mathbf V} =
\left[
\begin{array}{c c c c}
\uparrow  &  \uparrow  & \dots & \uparrow \\ 
{\bf v}_1  &  {\bf v}_2  &  \dots & {\bf v}_N \\
\downarrow &  \downarrow   & \dots & \downarrow 
\end{array}
\right]_{N \times N}, 
\quad 
{\bf D}_{\bf \lambda} =
\left[
\begin{array}{c c c c}
\lambda_1   &   0   & \ldots  &  0 \\
0     &   \lambda_2 & \ldots  &  0 \\
\vdots &      & \ddots  &  0 \\
0     &   0   & \ldots  &  \lambda_N
\end{array}
\right]_{N \times N}, 
\quad 
{\mathbf U} =
\left[
\begin{array}{c c c}
\longleftarrow  &  {\bf u}_1^T  &  \longrightarrow \\ 
\longleftarrow  &  {\bf u}_2^T  &  \longrightarrow \\ 
   & \vdots  &  \\ 
\longleftarrow  &  {\bf u}_N^T  &  \longrightarrow 
\end{array}
\right]_{N \times N}
\end{equation}

From eq.(\ref{eq:symW2}), (\ref{eq:symW3}) and (\ref{eq:matMVDU}) 

\begin{eqnarray} 
{\bf A}^k & = & {\bf V} {\bf D}_{\bf \lambda}^k {\bf U} \label{eq:symAk} \\
          & = & \sum_{i=1}^{N} \lambda_i^k {\bf v}_i {\bf u}_i^T  \label{eq:N1}
\end{eqnarray} 

From eq.(\ref{eq:genDLsolution}) and (\ref{eq:N1}), 

\begin{equation} \label{eq:solution} 
{\mathbf x}(k) = \sum_{i=1}^{N} \lambda_i^k {\bf v}_i {\bf u}_i^T {\mathbf x}(0) 
	+ \sum_{i=1}^{N} \sum_{j=1}^{k-1} \lambda_i^j {\bf v}_i {\bf u}_i^T {\bf b} + {\bf b}
\end{equation}

From the denominator of eq.(\ref{eq:cirA}), we define the interference vector, ${\bf J}(k)$, as follows 

\begin{equation} \label{eq:J_k} 
{\bf J}(k) = \Big( {\bf A} - {\bf D}_{\bf A} \Big) {\bf x}(k) + {\bf b}
\end{equation} 

where 

\begin{equation} \label{eq:diagW} 
{\bf D}_{\bf A} = 
\left[
\begin{array}{c c c c}
a_{11}   &   0   & \ldots  &  0 \\
0     &   a_{22} & \ldots  &  0 \\
\vdots &      & \ddots  &  0 \\
0     &   0   & \ldots  &  a_{NN}
\end{array} \right]_{N \times N}
\end{equation}

Defining $\mu_{u, i} = {\bf u}_i^T {\mathbf x}(0)$ and $\mu_{b, i} = {\bf u}_i^T {\bf b}$, 
eq.(\ref{eq:solution}) is rewritten as 

\begin{equation} \label{eq:solution_sum}
{\mathbf x}(k) = \sum_{i=1}^{N} \lambda_i^k {\bf v}_i \mu_{u, i} 
        + \sum_{i=1}^{N} \sum_{j=1}^{k-1} \lambda_i^j {\bf v}_i \mu_{b, i} + {\bf b}
\end{equation}

Let's show the (single or multiple) eigenvalue whose norm is the greatest among the eigenvalues 
as $\lambda_{max}$. Then, 
in eq.(\ref{eq:solution_sum}), the term related to the $\lambda_{max}$, 
whose norm is greater than 1 by assumption, dominates the sum.  
This is because a relatively small increase in $\lambda_{j}$ 
causes exponential increase as time step evolves, which is shown in the following: 
Let's define the following ratio 

\begin{equation} \label{eq:expDeltaL2}
\kappa(k)  = \frac{ (\lambda_{i})^{k} }{ ( \lambda_{i} + \Delta \lambda)^{k} }
\end{equation} 

where $\Delta \lambda$ represents the decrease (increase). 
The $\kappa(k)$ in eq.(\ref{eq:expDeltaL2}) is plotted in Figure 2 in \cite{Uykan09b} for some 
different $\Delta \lambda$ values. 
The Figure 2 in \cite{Uykan09b} implies that 
the term related to the $\lambda_{max}$ dominate 
the sum in eq.(\ref{eq:solution_sum}).  So, there exists a finite step number $k_T$ such that 

\begin{equation} \label{eq:x_sp}
{\bf x}(k) = \lambda_{max}^k {\bf v}_m \mu_{u, m}  
	+ \sum_{j=1}^{k-1} \lambda_{max}^j {\bf v}_m \mu_{b, m}, \quad \quad k \geq k_T 
\end{equation} 

where ${\bf v}_m$ is the eigenvalue corresponding to the $\lambda_{max}$.  
From eq.(\ref{eq:genDLsolution}), (\ref{eq:symAk}) and (\ref{eq:J_k}) 

\begin{eqnarray} \label{eq:J_k_mat}
{\bf J}(k) & = & {\bf V} {\bf D}_{\bf \lambda} {\bf U} {\bf x}(k) - {\bf D}_{\bf A} {\bf x}(k) + {\bf b} \\ 
	   & = & {\bf V} {\bf D}_{\bf \lambda}^{k+1} {\bf U} {\bf x}(0) + \sum_{j=1}^{k-1} {\bf V} {\bf D}_{\bf \lambda}^{j+1} {\bf U} {\bf b} 
	+ {\bf V} {\bf D}_{\bf \lambda} {\bf U} {\bf b} - {\bf D}_{\bf A} {\bf x}(k) + {\bf b}
\end{eqnarray}

Using eq.(\ref{eq:solution}) and (\ref{eq:J_k}) 

\begin{equation} \label{eq:Jk_solution} 
{\mathbf J}(k) = \sum_{i=1}^{N} \lambda_i^{k+1} {\bf v}_i {\bf u}_i^T {\mathbf x}(0) 
	+ \sum_{i=1}^{N} \sum_{j=1}^{k-1} \lambda_i^{j+1} {\bf v}_i {\bf u}_i^T {\bf b} 
	+ {\bf V} {\bf D} {\bf U} {\bf b} - {\bf D}_{\bf A} {\bf x}(k) + {\bf b} 
\end{equation} 

As in the steps from eq.(\ref{eq:solution_sum}) to eq.(\ref{eq:x_sp}), we 
see in eq.(\ref{eq:Jk_solution}) that 
the term related to the maximum eigenvalue $\lambda_{max}$ in magnitude, 
(which is greater than 1 by assumption), dominates the sum in (\ref{eq:Jk_solution}). 
In other words, comparing eq.(\ref{eq:solution_sum}), (\ref{eq:x_sp}) and 
(\ref{eq:Jk_solution}), we obtain that 

\begin{equation} \label{eq:Jk_sp}
{\bf J}(k) = \lambda_{max}^{k+1} {\bf v}_m \mu_{u, m} 
	+ \sum_{j=1}^{k-1} \lambda_{max}^{j+1} {\bf v}_m \mu_{b, m} 
 	- {\bf D}_{\bf A} {\bf x}(k), \quad \quad k \geq k_T 
\end{equation} 

where $\mu_{u, m} = {\bf u}_m^T {\mathbf x}(0)$ and $\mu_{b, m} = {\bf u}_m^T {\bf b}$ 
as defined in eq.(\ref{eq:x_sp}). 
Using eq.(\ref{eq:x_sp}) in (\ref{eq:Jk_sp}) 

\begin{equation} \label{eq:J_sp2} 
{\bf J}(k) = \{ \lambda_{max}{\bf I} -  {\bf D}_{\bf A} \} {\bf x}(k), \quad \quad k \geq k_T 
\end{equation}

In eq.(\ref{eq:Jk_sp}), we assume that $\mu_{u, m} \neq 0$, i.e., 
${\bf x}(0)$ is not completely perpendicular to ${\bf u}_{m}$ defined above.
That's why we say ``for ``almost'' any initial vector'' in the proposition phrase. 
However, this is something easy to check in advance. If it is the case, then
this can easily be overcome
by introducing a small random number to ${\bf x}(0)$ so that it's not completely perpendicular to 
the ${\bf u}_{m}$.  So, from eq.(\ref{eq:J_sp2}), 

\begin{equation} \label{eq:eta_maxL} 
\frac{x_{i}(k)}{J_{i}(k)}  =  \frac{1}{\lambda_{max} - a_{ii}}, \quad \quad k \geq k_T 
\end{equation} 

We conclude from eq.(\ref{eq:eta_maxL}) and from the "SIR" definition 
in eq.(\ref{eq:cirA}) that 

\begin{equation} \label{eq:cirA_sp}
{\theta_i}(k) = \frac{ a_{ii} x_{i}(k) }{ b_i + \sum_{j = 1, j \neq i}^{N} a_{ij} x_{j}(k) } = 
 	\frac{a_{ii}}{\lambda_{max} - a_{ii}}, 	\quad k \geq k_T, \quad  i=1, \dots, N,
\end{equation} 

where $\lambda_{max}$ is the (single or multiple) eigenvalue of ${\bf A}$ with the maximum norm and 
$k_T$ shows the finite time constant during which the $\lambda_{max}$ become dominant in the sum of 
(\ref{eq:solution_sum}), which completes the proof.

\end{proof}

\vspace{0.2cm}

\emph{Definition:} State-specific ultimate SIR value: In proposition 1, we showed that the SIR in (\ref{eq:cirA}) 
for every state in the autonomous discrete-time linear networks in eq.(\ref{eq:generalDLS}) converges to 
a constant value as step number goes to infinity.  We call this converged constant value 
as "ultimate SIR" and denote as ${\theta}^{ult}_i, \quad i=1,2, \dots, N$. 

\vspace{0.2cm}



Does the result in the proposition above have any practical meanings? Our answer is yes. 
In \cite{Uykan09a} and \cite{Uykan09b}, we present some (continuous-time and discrete-time) 
networks which are stabilized by the ultimate SIR as applied to the binary associative memory tasks 
as compared to the traditional Hopfield Neural Networks.
So, similar to the networks in \cite{Uykan09b}, the proposed autonomous network here is  

\begin{eqnarray}
{\mathbf x}(k+1) & = &  {\bf A} {\mathbf x}(k) 
             \delta ( {\bf \theta}(k) - {\bf \theta}^{ult} )  \label{eq:SAL-USIR2} \\
          {\mathbf y}(k) & = & sign( {\mathbf x}(k) )    \label{eq:SAL-USIR2n}
\end{eqnarray}

where ${\bf I}$ is the identity matrix, ${\bf A}$ is the system matrix, 
${\bf \theta}^{ult} = [{\theta}^{ult}_1 {\theta}^{ult}_2 \dots {\theta}^{ult}_N]^T$, and 
${\bf \theta}(k)= [ \theta_1(k) \theta_2(k) \dots \theta_N(k) ]^T $ 
is the SIR vector at step $k$ whose elements are given by eq.(\ref{eq:cirA}), 
the function $\delta(\cdot)$ gives 0 if and only if its argument vector is equal to zero vector, 
and gives 1 otherwise, and ${\mathbf y}(k)$ is the output of the network.

The proof of the proposition above shows that 
in the linear network of (\ref{eq:generalDLS}), 
the defined SIR in eq.(\ref{eq:cirA}) for state $i$  
converges to the constant SIR value in eq.
(\ref{eq:theta_const_general}),  
for any initial condition ${\bf x}(0)$ within a finite step number $k_T$.
It's seen that the linear networks of eq.(\ref{eq:generalDLS}) is nothing but the 
underlying network of the proposed network without the $\delta(\cdot)$ function. 
Since the ``SIR'' in eq.(\ref{eq:cirA}) exponentially approaches to the 
constant Ultimate ``SIR'' 
in eq.(\ref{eq:theta_const_general}), the delta function will stop the exponential 
increase once ${\bf \theta}(k) = {\bf \theta}^{ult}$, 
at which the system output reach its steady 
state response. So, the the presented network is stable. Furthermore, we note that 
the proposed networks in \cite{Uykan09a} and \cite{Uykan09b} which have been applied to 
the binary associative memory systems are special cases of the proposed network 
in eq.(\ref{eq:SAL-USIR2}) and (\ref{eq:SAL-USIR2n}). 
 
\begin{figure}[t!]
  \begin{center}
   \epsfxsize=24.0em    
\leavevmode\epsffile{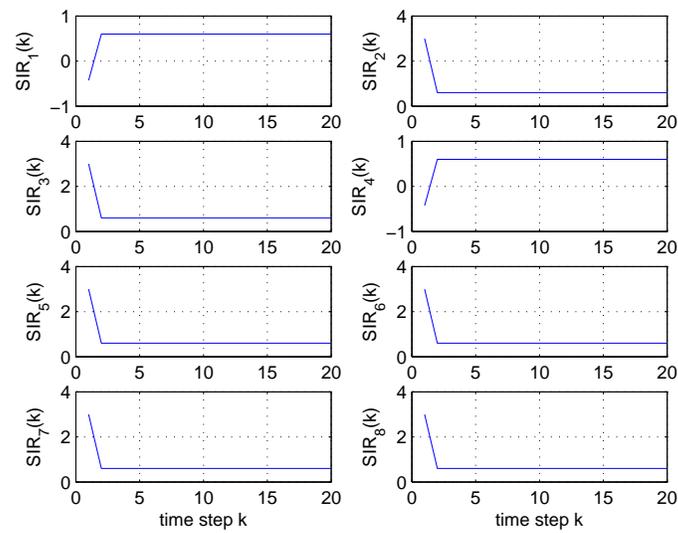}
   \vspace{-1em}        
  \end{center}
\begin{center} (a) \end{center}
  \begin{center}
   \epsfxsize=24.0em    
\leavevmode\epsffile{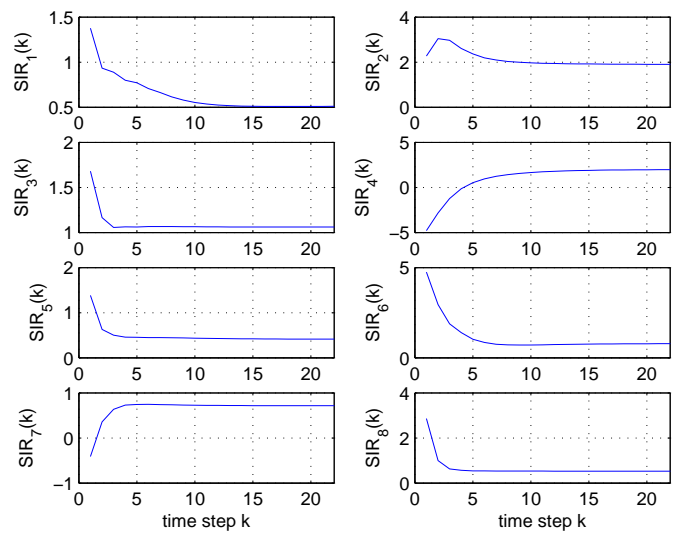}
   \vspace{-1em}        
  \end{center}
\begin{center} (b) \end{center}
 \caption
{ A sample evolution of SIR(k) with respect to time step for 8-dimensional system with 
(a) the weight matrix in \cite{Uykan09b} when the 
initial vector is 2-Hamming distance away from desired vector,
, and (b) random weight matrix. \label{fig:sirplot_n8_dHamming2_sample1}
}
\end{figure}

\section{Simulation Results  \label{Section:SimuResults} }

\begin{figure}[t!]
  \begin{center}
   \epsfxsize=24.0em    
\leavevmode\epsffile{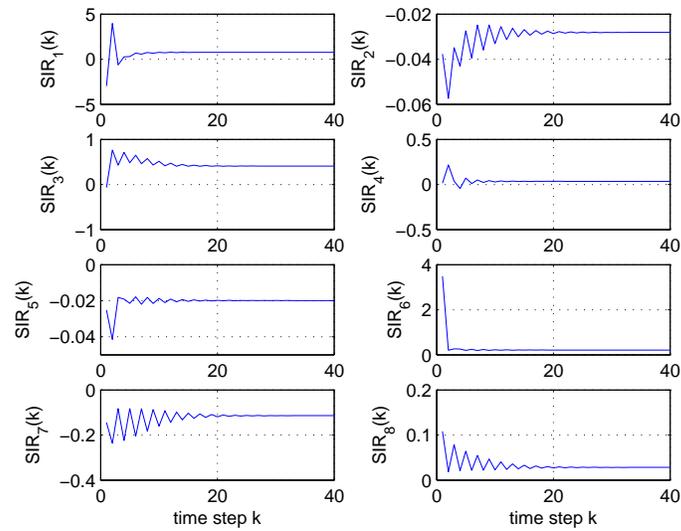}
   \vspace{-1em}        
  \end{center}
\begin{center} (a) \end{center}
  \begin{center}
   \epsfxsize=24.0em    
\leavevmode\epsffile{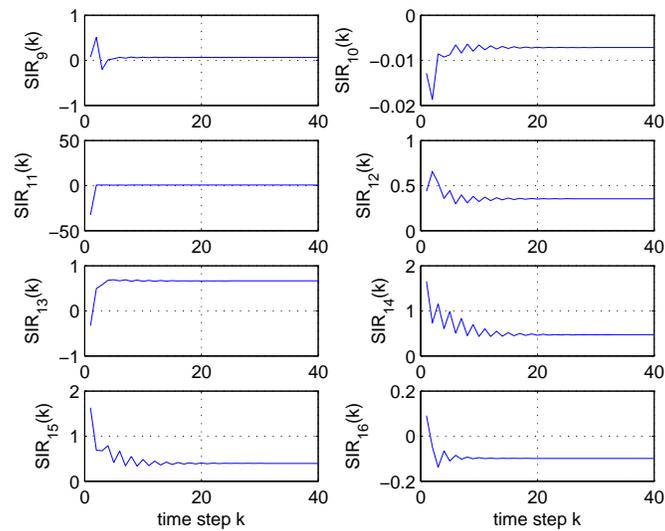}
   \vspace{-1em}        
  \end{center}
\begin{center} (b) \end{center}
\caption{ A sample evolution of SIR(k) with respect to time step for 16-dimensional system with 
random weight matrix, (a) SIRs from 1 to 8, (b) SIRs from 9 to 16. \label{fig:sirplot_randomW_n16_dHamming1_sample1a}
}
\end{figure}

In this section, we borrow the simple simulation examples in \cite{Uykan09b} with 8 and 16 neurons.  
We plot some samples of the evolution of the SIRs with respest to step number in the system 
of \cite{Uykan09b} whose weight matrix is presented in \cite{Uykan09b} as well as in 
the system of eq.(\ref{eq:generalDLS}) whose weight matrix is chosen randomly for 
illustration purpose.  
Figure \ref{fig:sirplot_n8_dHamming2_sample1}.a shows a sample of SIR evolution in the 8-neuron case in the 
example 1 of \cite{Uykan09b}, for initial 
condition $[-1 1 1 -1 -1 -1 -1 -1]$ while the desired vector is $[1 1 1 1 -1 -1 -1 -1]$ 
(Hamming distance is 2).  
Figure \ref{fig:sirplot_n8_dHamming2_sample1}.b shows a sample evolution of SIR in case of the system in 
eq.(\ref{eq:generalDLS}) with a random weight matrix. 
Some sample evolution plots for the 16-neuron case are shown in 
Figure \ref{fig:sirplot_randomW_n16_dHamming1_sample1a} for the system of eq.(\ref{eq:generalDLS}) 
with random weight matrix.  
The figures \ref{fig:sirplot_n8_dHamming2_sample1} and 
\ref{fig:sirplot_randomW_n16_dHamming1_sample1a} show that the SIRs converge to their ultimate values and 
confirm the results in previous section.

\section{Conclusions  \label{Section:CONCLUSIONS}}

In this letter, we improve the SIR results in \cite{Uykan09b} 
by relaxing the symmetry assumption and taking also the noise term into account. 
We examine the following discrete-time autonomous 
linear system: ${\mathbf x}(k+1) = {\mathbf A} {\mathbf x}(k) + {\mathbf b}$, where 
${\mathbf A}$ is any real square matrix with linearly independent eigenvectors
whose largest eigenvalue is real and its norm is larger than 1, and 
vector ${\mathbf b}$ is constant. 
Using the same ``SIR'' concept as in \cite{Uykan09a} and \cite{Uykan09b}, 
we show that the ultimate ``SIR'' is equal to $\frac{a_{ii}}{\lambda_{max} - a_{ii}}$, $i=1, 2, \dots, N$, 
where $N$ is the number of states, $\lambda_{max}$ is the eigenvalue with maximum norm and 
$a_{ii}$ is the diagonal elements of matrix ${\bf A}$.


%



%
\section*{Acknowledgments} 

This work was supported in part by Academy of Finland and Research Foundation (Tukis\"{a}\"{a}ti\"{o}) of Helsinki University of Technology, Finland.

\nocite{*}
\bibliographystyle{IEEE}

%




\end{document}